\documentclass[a4paper,twocolumn,%tightenlines,
english,aps,pre,floatfix,groupedaddress,showpacs,nofootinbib]{revtex4-1}
\usepackage[T1]{fontenc}
\usepackage[latin1]{inputenc}
\usepackage{amsmath}
\usepackage{babel}
\usepackage{graphics}
\usepackage{amssymb}
\usepackage{mathrsfs}
\usepackage{dcolumn}
\usepackage{epstopdf}


\begin{document}
\title
{Driven flow with exclusion and spin-dependent transport in graphenelike
structures} 
\author {S. L. A. \surname{de Queiroz}}
\email{sldq@if.ufrj.br}
\affiliation{Instituto de F\'\i sica, Universidade Federal do
Rio de Janeiro, Caixa Postal 68528, 21941-972
Rio de Janeiro RJ, Brazil}
\author {R. B. \surname{Stinchcombe}}
\email{Robin.Stinchcombe@physics.ox.ac.uk}
\affiliation{Rudolf Peierls Centre for Theoretical Physics, University of
Oxford, 1 Keble Road, Oxford OX1 3NP, United Kingdom}

\date{\today}

\begin{abstract} 
We present  a simplified description for spin-dependent electronic transport in 
honeycomb-lattice structures with spin-orbit interactions,
using generalizations of the stochastic non-equilibrium model known as 
the totally asymmetric simple exclusion process. 
Mean field theory and numerical simulations are used to study
currents, density profiles and current polarization in quasi- one dimensional
systems with open boundaries, and externally-imposed particle injection ($\alpha$)
and ejection ($\beta$) rates. We investigate the influence of allowing for double site
 occupancy,
according to Pauli's exclusion principle, on the behavior of the quantities of
interest. We find that double occupancy shows strong signatures for 
specific combinations of rates, namely high $\alpha$ and low $\beta$, but
otherwise its effects are quantitatively suppressed. Comments are made on the possible
relevance of the present results to experiments on suitably doped graphenelike
structures.   
\end{abstract}
%\pacs{05.40.-a, 02.50.-r, 72.80.Vp, 73.23.-b}
%05.40.-a Fluctuation phenomena, random processes, noise, and Brownian motion
%02.50.-r Probability theory, stochastic processes, and statistics
%72.80.Vp Electronic transport in graphene
%73.23.-b Electronic transport in mesoscopic systems
\maketitle
%\tightenlines
 
\section{Introduction} 
\label{intro} 

In this paper we consider a simplified model for spin-dependent
electronic transport in honeycomb-lattice structures with 
spin-orbit (SO) interactions. Suitable generalizations of
the totally asymmetric simple exclusion process (TASEP) are
applied to such systems. This extends previous work which dealt
with steady-state properties~\cite{hex13} and dynamics~\cite{hex15}
of the most basic implementation of the TASEP on 
honeycomb-lattice geometries.

The TASEP, in its one-dimensional (1D) version, already exhibits
many non-trivial properties because of its collective 
character~\cite{derr98,sch00,mukamel,derr93,rbs01,be07,cmz11}.
It has been used, often with adaptations, to model a broad range
of non-equilibrium physical phenomena, from the macroscopic level such
as highway traffic~\cite{sz95} to the microscopic, including sequence
alignment in computational biology~\cite{rb02} 
and current shot noise in quantum-dot chains~\cite{kvo10}.

In the time  evolution  of the 1D TASEP,
the particle number $n_\ell$ at lattice site $\ell$ can be $0$ or $1$, 
and the forward hopping of particles is only to an empty adjacent site. 
In addition to the stochastic character provided by random selection of site 
occupation 
update~\cite{rsss98,dqrbs08}, the instantaneous current ${\hat J}_{\ell\,\ell+1}$ 
across the bond from $\ell$ to $\ell+1$ depends also on the stochastic 
attempt rate, or bond (transmissivity) rate, $t_\ell$, associated with it. 
Thus,  
\begin{equation}
{\hat J}_{\ell\,\ell+1}= \begin{cases}{n_\ell (1-n_{\ell+1})\quad {\rm with\ probability}\ t_\ell}\cr
{0\qquad\qquad\qquad {\rm with\ probability}\ 1-t_\ell\ .}
\end{cases}
\label{eq:jinst}
\end{equation}
In Ref.~\onlinecite{kvo10} it was argued that the ingredients of 1D TASEP
are expected to be physically present in the description of electronic
transport on a quantum-dot chain; namely,
the directional bias would be provided by an external voltage difference
imposed at the ends of the system, and the exclusion effect by on-site Coulomb 
blockade. Following similar lines, the present work with its 
emphasis on honeycomb structures is partly motivated
by recent progress in the physics of graphene and its quasi-1D
realizations, such as nanotubes and nanoribbons~\cite{RMP,peres,ml10}.
 Being a classical model,  the TASEP  does not incorporate quantum 
interference effects which play an important role in electronic transport.
However, when considering transport in carbon allotropes under an applied 
bias the lattice topology affects how currents combine, and how they are 
microscopically located, whether classical or quantum. These features show 
up in the model we treat by such effects as the sublattice structure seen
in steady-state currents and densities for the hexagonal 
lattice~\cite{hex13,hex15}. 

Here we focus on modeling the behavior of spin polarization~\cite{zfds04}
in graphene-like quasi--1D geometries, in the presence 
of SO couplings~\cite{smr15}. The spin-flipping
character of SO interactions  is represented, e.g., in a tight-binding description,
by a non-diagonal $2 \times 2$ matrix
in the space of eigenfunctions of the electron's spin 
$\sigma_z$~\cite{ando89,ez87,e95,ando00,aso04,dq15}.  Although the 
effective strength of intrinsic SO coupling in graphene is estimated to be $25-50$ $\mu$eV~\cite{kgf10}, much smaller than the nearest-neighbor hopping
$\gamma_0=2.8$ eV~\cite{RMP}, doping with suitable impurities can result
in samples where SO interactions are more significant in specific neighborhoods
next to impurity locations~\cite{rigo09,nat13,dope3,dope4}. 

In Sec.~\ref{sec:tasep-theo} we briefly recall basic features of 
the spin-independent TASEP
model used in Refs.~\onlinecite{hex13,hex15}, and  outline the 
adaptations and approximations
here added to the model,  in order to describe spin polarization, 
SO interactions,  and spin-dependent currents. A mean field theoretical
description is given for the problem.
The corresponding numerical  tests are given in Sec.~\ref{sec:tasep-num}. 
Section~\ref{sec:conc} is devoted to
discussions and conclusions.

\section{TASEP model: theory}
\label{sec:tasep-theo}

\subsection{Introduction}
\label{theo-intro}

We only consider cases where mean flow direction is 
parallel to one of the lattice directions, and bond rates are independent of coordinate 
transverse to the flow direction. These configurations have no bonds orthogonal to the 
mean flow direction; thus they fall easily within the generalized TASEP description to be 
used, where each bond is to have a definite directionality, compatible with that of 
average flow. Furthermore, for simplicity we always use periodic boundary 
conditions (PBC) across. So, in the terminology of quasi 1D carbon nanotubes 
(CNT) and nanoribbons (CNR)~\cite{RMP}, the structures to be discussed  
correspond to zigzag CNTs. 
Most experimental studies, as well as many theoretical ones, deal with 
impurities on CNRs. However, edge effects play an important role in the 
energetics of favored defect locations for the latter type of system. Since here we 
shall not attempt detailed numerical comparisons to experimental data,
our choice of considering only nanotube geometries where we need not account 
for this sort of  positional preference inhomogeneity, is justified on grounds
of keeping the number of relevant parameters to a minimum.

The structures studied here have  an integer number of elementary cells 
(one bond preceding a full hexagon) along the mean flow direction, see
Refs.~\onlinecite{hex13,hex15}. Also, we have to expect a two-sublattice
character in general~\cite{hex13}.

Open boundary conditions are used at both ends of the strip with the 
associated, externally imposed, injection and ejection (attempt) rates 
$\alpha$ and $\beta$~\cite{derr98,derr93}. For all internal bonds $\ell$ 
we take their transmissivity rates, defined in Eq.~(\ref{eq:jinst}), to 
be $t_\ell \equiv~1$.

Thus a nanotube with $N_r$ elementary cells 
parallel to the flow direction, and $N_w$ transversally, has 
$N_s=N_w \times (4N_r+1)$ sites and $N_b=N_w \times (6N_r+2)$ bonds (including
the injection and ejection ones).   

In the TASEP context, the simplest way to simulate the effects 
associated with SO couplings in doped systems is by assigning a quenched 
random distribution of {\em spin-flipping sites} (substitutional 
impurities) to an otherwise pure sample, with the following rule: every 
time a particle passes through such a site the $z$ component of its spin 
$1/2$ will change sign, with probability $\pi_f$ (or remain the same, with 
probability $1-\pi_f$).  In what follows, we shall always take $\pi_f=1$ for 
simplicity. All the original TASEP rules are kept except that, regarding 
exclusion effects, the physical properties of the problem under 
investigation immediately suggest two plausible alternatives:

\noindent (A)\ Keep the maximum occupation per site $n_\ell=1$ as in 
the original formulation, or

\noindent (B)\ Allow two particles simultaneously on the same 
site, provided that their spins are opposite (thus mimicking Pauli's 
principle).

In model (A), polarization and global current (and density) aspects are 
decoupled. Thus, although the evolution of spin polarization along the 
system shows interesting and nontrivial features, overall currents and 
total (spin-up plus spin-down) density profiles will be the same as in 
the spinless cases studied in Refs.~\onlinecite{hex13,hex15}.  On the 
other hand, we will see that model (B) exhibits some rather intricate 
interplay between spin and real-space degrees of freedom.

\subsection{Mean field description}
\label{sec:theo-mf}

\subsubsection{Impurity-induced spin polarization decay}
\label{sec:em}

Initially we give a simplified approach to describe the behavior
of spin polarization, by focusing on the path followed by a single particle
traveling along the system, in the course of which spin-flipping 
events may occur.
With $x_i \ll 1$ being the concentration of spin-flipping (impurity) 
sites, the TASEP directionality rules imply that, for a system with 
$N_r$ rings along the flow direction, each particle flowing through the 
nanostructure will have to go through exactly $N=4N_r +1$ sites. Assume 
impurities to be uniformly distributed along the system according to a
grand-canonical distribution with mean $x_i$; 
%%%%%%%%%%%%%%%% 1st referee, point (1) %%%%%%%%%%%%%%%%%%%
replace each slice of the nanotube across the average flow direction
with an "effective" site  ($N$ in all); and take $x_i$ as the probability 
of stepping on an impurity at each "effective" site. So we are replacing 
the actual realizations of random discrete spin-flipping sites by a 
homogeneous "effective medium"  with a probability per slice $x_i$ of the
particle's spin being flipped.
%%%%%%%%%%%%%%%%%%%%%%%%%%%%%%%%%%%%%%%%%%%%%%%%%%%%%%%%%%%%%%%%
Recalling that 
the particle's spin upon  exiting at
the ejection point will depend only on whether it has met an even or 
odd number of impurities, elementary probabilistic considerations allow 
one to work out the exact probability distribution for the current 
polarization at any cross section (take, for simplicity, a 
fully-polarized injected current).  
The spin fraction of the average exiting current with plus or minus
spin turns out to be, for $N \gg 1$,
\begin{equation}
P_\pm= \frac{1}{2} \left[ 1\pm(1-2x_i)^N\right]\ ,
\label{eq:binom1}
\end{equation}
whence the (ensemble-averaged) exit polarization is
\begin{equation}
\langle {\cal P}(x_i)\rangle \propto \exp (-N/N_0)\ , \quad N_0=-[\ln(1-2x_i)]^{-1}\ .
\label{eq:binom2}
\end{equation}
Such an effective-medium approach of 
course neglects all correlations between particle occupation at 
neighboring sites and the corresponding local currents, recall 
Eq.~(\ref{eq:jinst}); 
its predictions are also independent of whether model (A) or (B) is
adopted. Furthermore, the results for normalized 
polarizations will depend only on $x_i$ and on the position $\ell$ ($1 
\leq \ell \leq 4N_r+1$) of the cross-section under consideration, and 
{\em not}, e.g., on the value of the ($\alpha$- and $\beta$- dependent) 
steady-state current through the system. In Sec.~\ref{sec:tasep-num} we 
test the predictions of this simplified description against the results 
of numerical simulations.

\subsubsection{Currents and density profiles}
\label{sec:mf-general}

The mean field description of total currents and density profiles in 
model (A) is identical to that for spinless models, and is given at length
in Refs.~\onlinecite{hex13,hex15}.

In model (B) there are four mutually exclusive possibilites for occupation of a 
site: vacant, one particle with plus spin, one particle with negative spin,
and two particles with opposite spins. Their associated probabilities for 
site $\ell$ are denoted respectively by $p_\ell^0$, $p_\ell^+$, $p_\ell^-$, 
$p_\ell^{+-}$, with $p_\ell^0+p_\ell^++p_\ell^-+p_\ell^{+-}=1$.

We also use here corresponding state indicator variables ${\hat p}_\ell^0$, ${\hat p}_\ell^+$, 
${\hat p}_\ell^-$, ${\hat p}_\ell^{+-}$, such that, for example, ${\hat p}_\ell^+$ can be one or 
zero, respectively specifying that site $\ell$ is occupied or not by a spin $+$ particle. Then $p_\ell^+$
is the average of ${\hat p}_\ell^+$; and so on .

Ignoring, until stated otherwise, the possible effect of impurities, the possible 
updates of bonds linking, say, sites $\ell$ and $\ell +1$ contributing to 
the instantaneous plus-spin current ${\hat J}^+$
through transfer across of a $+$ spin particle
have initial configurations with the following indicators: ${\hat p}_\ell^+{\hat p}_{\ell+1}^0$,
${\hat p}_\ell^+{\hat p}_{\ell+1}^-$, ${\hat p}_\ell^{+-}{\hat p}_{\ell+1}^-$, 
${\hat p}_\ell^{+-}{\hat p}_{\ell+1}^0$.

Those for ${\hat J}^-$ have a similar set but with $+$ and $-$ superscripts interchanged.

From the bond update details given at the beginning of Sec.~\ref{sec:tasep-num},
unit bond rates are associated with the first three configurations listed for each
of ${\hat J}^+$ and ${\hat J}^-$, while the rate is $\frac{1}{2}$ for the last (shared)
configuration. Consequently, in place of Eq.~(\ref{eq:jinst}), for any configuration of
the system with model (B) the currents on bond $\ell$, $\ell+1$ are given (exactly)
in terms of the indicator variables [$\,$collectively denoted as $\{{\hat p}\}\,$] by:
\begin{eqnarray}
{\hat J}_{\ell\,\ell+1}^+& = &{\hat p}_\ell^+{\hat p}_{\ell+1}^0 +
\frac{1}{2}{\hat p}_\ell^{+-}{\hat p}_{\ell+1}^0+{\hat p}_\ell^+{\hat p}_{\ell+1}^-+
{\hat p}_\ell^{+-}{\hat p}_{\ell+1}^- \nonumber \\
&  & \equiv C_{\ell\,\ell+1}^+({\{\hat p}\})\ ; 
\label{eq:jindp}
\end{eqnarray}
\begin{eqnarray}
{\hat J}_{\ell\,\ell+1}^-& = &{\hat p}_\ell^-{\hat p}_{\ell+1}^0 +
\frac{1}{2}{\hat p}_\ell^{+-}{\hat p}_{\ell+1}^0+{\hat p}_\ell^-{\hat p}_{\ell+1}^++
{\hat p}_\ell^{+-}{\hat p}_{\ell+1}^+ \nonumber \\
&  & \equiv C_{\ell\,\ell+1}^-({\{\hat p}\})\ .
\label{eq:jindm}
\end{eqnarray}
The currents ${\hat J}_{\ell\,\ell+1}^+$, ${\hat J}_{\ell\,\ell+1}^-$ have no terms with factors ${\hat p}_\ell^0$
or ${\hat p}_{\ell+1}^{+-}$, for obvious physical reasons. With ${\hat I}$ denoting the
identity indicator, that means that they can,
for example, be multiplied by $({\hat I}-{\hat p}_{\ell+1}^{+-})$, to give new forms
${\hat J}_{\ell\,\ell+1}^\pm =C_{\ell\,\ell+1}^{\pm\,\prime}=C_{\ell\,\ell+1}^\pm(
{\hat I}-{\hat p}_{\ell+1}^{+-})$
which are still exact.

The most direct and obvious mean field approximation for the mean currents $J^+$, $J^-$
is obtained by replacing each ${\hat p}$ in the $C^+$, $C^-$ of Eqs.~(\ref{eq:jindp})--~(\ref{eq:jindm})
by its average $p$. The resulting form proves to be entirely adequate for most purposes in this
investigation. But equally well one could have made the mean field replacement using the
equivalent forms $C^{+\,\prime}$, $C^{-\,\prime}$, which results in a different mean field description.

The latter description is slightly more complicated but it might be expected
to better capture the physics of the process in the high density regions,
where large $p_{\ell+1}^{+-}$ suppresses the current. So it will be used 
in Sec.~\ref{sec:num-cnt} for one such case. Apart from there
we use mean field approximations with mean currents 
$J_{\ell\,\ell+1}^\pm=C_{\ell\,\ell+1}^\pm(\{p\})$.

The original mean field theory for the TASEP chain assumes approximate factorization 
of correlation functions. In the same spirit we will approximate $p_\ell^{+-}$ by
$p_\ell^+\,p_\ell^-$. It will be seen that this gives completeness to the set of
steady state mean field equations that arises from conservation of the $+$ and $-$ spin
currents $J^+$, $J^-$, at each vertex including boundary vertices $\ell=0$, $L$ in
the case of open boundary conditions. With this approximation, the mean number
 of $+$ ($R_\ell$) or $-$ ($S_\ell$) spin particles at site $\ell$ are respectively:
\begin{equation}
R_\ell=p_\ell^+(1+p_\ell^-)\quad ;\quad S_\ell=p_\ell^-(1+p_\ell^+)\ .
\label{eq:rs}
\end{equation} 

Then, using also $p_\ell^0+p_\ell^++p_\ell^-+p_\ell^{+-}=1$, the mean field
equations for the average currents become
\begin{eqnarray}
J_{\ell\,\ell+1}^+& = &{\hat p}_\ell^+\left(1+\frac{1}{2}p_\ell^-\right)\left[1-(p_{\ell+1}^++
p_{\ell+1}^-+p_{\ell+1}^+p_{\ell+1}^-)\right] \nonumber\\
&  & + p_\ell^+(1+p_\ell^-)p_{\ell+1}^- \equiv f(p_\ell^+,p_\ell^-,p_{\ell+1}^+,
p_{\ell+1}^-)\ ;
\label{eq:jp}
\end{eqnarray}
\begin{eqnarray}
J_{\ell\,\ell+1}^-& = &p_\ell^-\left(1+\frac{1}{2}p_\ell^+\right)\left[1-(p_{\ell+1}^++
p_{\ell+1}^-+p_{\ell+1}^+p_{\ell+1}^-)\right] \nonumber\\
&  & + p_\ell^-(1+p_\ell^+)p_{\ell+1}^+ \equiv f(p_\ell^-,p_\ell^+,p_{\ell+1}^-,
p_{\ell+1}^+)\ .
\label{eq:jm}
\end{eqnarray}
Similarly, with $+$ spin particles injected at rate $\alpha^+$ at the left boundary
site $\ell=0$ the mean current $J^+$ entering there is
\begin{equation}
J_0^+ = \alpha^+\left[1-p_0^+(1+p_0^-)\right]\ .
\label{eq:jp0}
\end{equation}
Likewise, with ejection rate $\beta^+$ for $+$ spin particles at the right boundary 
site $\ell=L$, the mean current $J^+$ leaving there is
\begin{equation}
J_L^+ =\beta^+p_L^+(1+p_L^-)\ .
\label{eq:jpL}
\end{equation}
The corresponding boundary currents for $-$ spin particles satisfy corresponding
equations in which the $+$ and $-$ signs are interchanged.

The study of steady state properties involves relating the currents $J^+$, $J^-$
and the density profiles $R_\ell$, $S_\ell$ to the injection and ejection rates
$\alpha^+$, $\alpha^-$, $\beta^+$, $\beta^-$. In general this involves the use
of profile maps resulting from current conservation at each lattice vertex.

\subsubsection{Model (B) on linear chain (without impurities)}
\label{sec:bchainpure}

As will be seen below, for model (B) on the nanotube the maps are quite complex, 
being two-stage maps of four sets of variables ($p_\ell^+$ and $p_\ell^-$ for
two sublattices), so we first go to the simpler and more transparent case of
model (B) on the linear chain, still without impurities.

There, in the steady state all bonds $\ell$, $\ell+1$ carry the same $J^+$ and the 
same $J^-$, which must also equal the injected and ejected currents. For $0 \leq \ell
\leq L-1$, using Eqs.~(\ref{eq:jp}) and~(\ref{eq:jm}),
\begin{equation} 
f(p_\ell^+,p_\ell^-,p_{\ell+1}^+,p_{\ell+1}^-)=J_0^+=J_L^+\ ;
\label{eq:contp}
\end{equation}
\begin{equation} 
f(p_\ell^-,p_\ell^+,p_{\ell+1}^-,p_{\ell+1}^+)=J_0^-=J_L^-\ ,
\label{eq:contm}
\end{equation}
where $J_0^+$, $J_L^+$ are given in Eqs.~(\ref{eq:jp0}) and~(\ref{eq:jpL})
(similarly for $J_0^-$, $J_L^-$).

Eqs.~(\ref{eq:contp}),~(\ref{eq:contm}) constitute the "one-stage" map giving, in 
principle, 
for specified currents, $p_{\ell+1}^+$ and $p_{\ell+1}^-$ in terms of $p_\ell^+$
and $p_\ell^-$ and finally, using the detailed forms of the injection and ejection 
currents, all $p_\ell^+$, $p_\ell^-$ and currents in terms of the boundary rates.
This description properly handles, through the specific forms of current in 
Eqs.~(\ref{eq:jp}),~(\ref{eq:jm}) the new effects of the conditional double occupancy 
in model (B). Furthermore, the description is clearly complete (with the use of the
approximation $p^{+-}=p^+ p^-$). 

By exploiting the symmetries and relative simplicity of the dependences on
$p_\ell^+$, $p_\ell^-$, $p_{\ell+1}^+$, $p_{\ell+1}^-$ of the combinations
$J_{\ell\,\ell+1}^+ \pm J_{\ell\,\ell+1}^-$ it is possible to obtain explicit
functional forms for  $p_{\ell+1}^+$, $p_{\ell+1}^-$ in terms of
$p_\ell^+$, $p_\ell^-$, and the constant values of $J^+$, $J^-$:
\begin{equation}
p_{\ell+1}^+= \frac{1}{2A}\left[-(B+CA)\pm \sqrt{(B+CA)^2-4A(Cp_\ell^--D)}\right]
\label{eq:plp1p}
\end{equation}
\begin{equation}
p_{\ell+1}^-=p_{\ell+1}^++C\ ,
\label{eq:plp1m}
\end{equation}
where
\begin{eqnarray}
A=p_\ell^++p_\ell^- +p_\ell^+p_\ell^- \ ;\ B=p_\ell^+ + p_\ell^-\ ;\nonumber \\
 D=A-(J^++J^-)\ ;\ C=\frac{J^+-J^-}{p_\ell^+p_\ell^-}\ .
\label{eq:abcd}
\end{eqnarray}
This gives the explicit two-variable profile map for model (B) on the linear chain.

We next consider the possible fixed points of the map. There can be two real 
(physical) ones, at $(p_\ell^+,p_\ell^-)=(p_<^+,p_<^-)$ and $(p_>^+,p_>^-)$, 
the first ("lower") one having entries lower than those in the other ("upper") one,
and they correspond to unstable (repulsive) and stable (attractive) ones, respectively,
in the forward mapping $\ell \to \ell+1$. 

So, e.g., starting with $p^+$ and $p^-$
between the fixed points and very close to the lower one, at first many iteration 
steps leave  $(p^+,p^-)$ close to the starting value before it rapidly moves away and 
homes in on the attractive fixed point. So the associated profile $(p_\ell^+,p_\ell^-)$
has $p_\ell^+$ and $p_\ell^-$ both monotonically increasing with $\ell$ and each 
qualitatively similar to the low-current-phase form $a+b\tanh (\phi +\ell \theta)$
of the mean field TASEP chain, with $a-b=p_<$, $a+b=p_>$ for each of
$p_\ell^+$ and $p_\ell^-$.

For the other possibilities (two coincident physical fixed points or none) the 
possible profiles are again qualitatively similar to those of the TASEP at its
critical point or in its maximal current phase (i.e., with $-\tan$ replacing 
$\tanh$). So for model (B) on the chain without impurities the mean field phase 
diagram and density profiles in the various phases are like those of the TASEP
chain.

For a quantitative example we proceed next from the full formalism of 
Eqs.~(\ref{eq:plp1p})--(\ref{eq:abcd}) to the case with boundary conditions,
such that the map has a fixed point at $p_\ell^+=p_\ell^- \equiv x^\ast$, corresponding 
to equal and level density profiles for spin up and spin down particles (hence, 
unpolarized). From the results above, this is only possible if
\begin{equation}
J^+=J^-= x^\ast -\frac{1}{2}x^{\ast 2}- x^{\ast 3}-\frac{1}{2}x^{\ast 4} \ .
\label{eq:jvsxast}
\end{equation}
The level profiles will normally extend to one or the other boundary,
depending on the relative sizes of the injection and ejection rates.
Then we may use the relationship of $J^+$, $J^-$ to the rates at that
boundary. For example, when that boundary is the injection site the above
analysis shows that for equal level profiles $\alpha^+=\alpha^-$ is needed, 
and then
\begin{equation}
J^+=J^-=\alpha\left[1-x^\ast(1+x^\ast)\right]\ .
\label{eq:jvsalpha}
\end{equation}
The corresponding level spin up and spin down particle densities are then
\begin{equation}
R=S=x^\ast(1+x^\ast)\ ,
\label{eq:rsxast}
\end{equation}
with no net polarization. 

Eq.~(\ref{eq:jvsxast}) also allows the extraction of the maximal current $J_m$ for level
profiles, and shows that it occurs when the two fixed points become coincident. This 
is because the function on the right hand side of Eq.~(\ref{eq:jvsxast}) has a single
maximum in the physical region $x^\ast \geq 0$. It is then the $J^+=J^-$ value at
which the two solutions (fixed points) come together. One gets 
$J_m^+=J_m^-=0.243207\dots$ for $x_m^\ast=0.39817 \dots$, so $R_m=S_m=0.55671 \dots$. 
These predictions are compared to simulational results in Sec.~\ref{sec:num-chain} below.

Finally, concerning model (B) on the pure chain we note that with fully polarized 
injection, e.g., $\alpha^+=1$, $\alpha^-=0$  no double occupancy occurs at any site,
so all properties become identical to those of the standard TASEP.

\subsubsection{Model (B) on nanotube (without impurities)}
\label{sec:bcntpure}

As in previous studies~\cite{hex13,hex15}, what follows for model (B) on the nanotube
concerns situations with azimuthal symmetry. The "top" open boundary of the
hexagonal nanotube is taken to be the ring of sites all with $\ell=0$, 
all having the same injection rates $(\alpha^+,\alpha^-)$; similarly with the
ejection sites, at $\ell=L$ at the other end of the tube. So $\ell$ indicates
site position down the tube and no site azimuth coordinate will be needed for
the steady state properties here discussed.

For the TASEP on the nanotube it is necessary to distinguish two sublattices, 
"even" and "odd". At each interior site of the even sublattice 
 (having even label) two bonds are incident 
from above and one leaves below, and vice versa for all odd sublattice sites; 
for a pictorial representation see Fig.~1 of Ref.~\onlinecite{hex15}.
We take $L$ even ($L=2M$) so that injection and ejection are on the same 
(even) sublattice.

Because of the branching and recombination of bonds referred to above, in the steady
state each bond $(2\ell,2\ell+1)$ (for all $\ell$) carries the same $(J^+,J^-)$
which is twice that for each bond $(2\ell+1,2\ell+2)$ (all $\ell$) and the same as
the injection $(J_0^+,J_0^-)$ and ejection $(J_L^+,J_L^-)$. That is, 
for $\sigma=\pm$,
\begin{equation}
J^\sigma_{2\ell\,2\ell+1}=2J^\sigma_{2\ell+1\,2\ell+2}= J^\sigma_0=J^\sigma_L\ ,
\label{eq:jejo}
\end{equation}
for any $\ell$ in $[0,M-1]$.

The $J^\sigma$'s here are as given in terms of $\{p_\ell^+,p_\ell^-\}$ by
Eqs.~(\ref{eq:jp})--(\ref{eq:jpL}). The current-balance Eqs.~(\ref{eq:jejo})
above provide the mapping relationships  between probability
variables at successive positions $\ell$, in principle enough to find all
mean field profiles and currents in terms of boundary rates. But successive
sites lie on different sublattices so, as in previous studies~\cite{hex13,hex15}
any, even qualitative, connection with analytic functions requires a two-stage
map between adjacent sites on the same sublattice, via one on the other sublattice.
The quadratic dependence of bond currents on $p$ variables of both sites they
link, coming from the conditional double occupancy, is a further complicating
feature, making most further progress purely numerical.

Nevertheless more analytic progress is possible concerning the fixed points of
the two-stage mapping for a given sublattice, whose connection with level
portions of particle density profiles associated with low current phases, and 
with maximal current aspects have been exploited in Sec.~\ref{sec:bchainpure} 
above. We next consider these, using the example of unpolarized cases.

For fully unpolarized systems, e.g., arising from $\alpha^+=\alpha^-=\alpha$,
$\beta^+=\beta^-=\beta$, we have $p_\ell^+=p_\ell^-\equiv p_\ell$, and
$J^+=J^-$ on each sublattice. The fixed points of the two-step map then
correspond to having all $p_\ell$ the same on each sublattice, i.e.,
\begin{equation}
p_{2\ell}=u\ ,\quad p_{2\ell+1}=v\ \qquad ({\rm all}\ \ell)\ ,
\label{eq:cntlevel}
\end{equation} 
with $u$ and $v$ the level values of the probabilities at the sites of the two
sublattices.

Then, from Eq.~(\ref{eq:jejo}) the two-stage map from a site on the even sublattice
to a forward adjacent site (on the odd sublattice), and then from there
to a further forward adjacent site on the even sublattice is
\begin{equation}
f(u,u,v,v)=J_0=J_L\ ;\qquad f(v,v,u,u)=\frac{1}{2}J_0\ .
\label{eq:cntmap}
\end{equation}
Numerical methods can readily provide solutions of these equations for $u$, $v$
for a range of specified values of $J_0$, from zero to a cutoff at $J_0 \approx 0.332$.
For small $J_0$, the behavior of $u$ and $v$ is given by
\begin{equation}
u=J_0+{\cal O}(J_0^3)\ ;\qquad v=\frac{J_0}{2} + \frac{5}{8}J_0^2+ {\cal O}(J_0^3)\ .
\label{eq:rsvsjo}
\end{equation}
The resulting $u$, $v$ can be inserted into the injection and ejection current 
forms, Eqs.~(\ref{eq:jp0}) and~(\ref{eq:jpL}) to obtain corresponding boundary rates.
Such results, and corresponding level sublattice particle densities $U=u(1+u)$,
$V=v(1+v)$ allow comparisons with results from steady state simulations.   

Finally in this Subsection, still for the fully
unpolarized case, we give results for the dividing line on 
the $(\alpha,\beta)$ phase diagram, between current 
independent of $\alpha$ and independent of $\beta$. 
By analogy with the standard 1D TASEP~\cite{derr98,sch00,mukamel,derr93,rbs01}
this could possibly be the signature of a coexistence line, 
though we shall not investigate such a connection here. 

The calculation involves the full 
range of possible site average occupations. Both forms of mean field theory
outlined in Sec.~\ref{sec:mf-general} were used. That from the $C^{\pm}$ gives
the line equation as
\begin{equation}
\alpha\,\frac{\left(1-\frac{\alpha}{2}\right)}{\left(1+\frac{\alpha}{2}\right)}=
2\beta\,\frac{\left(1-\beta\right)}{2-\beta}\ .
\label{eq:beta(a)_th}
\end{equation}
A more complicated equation results from using the mean field theory from the
$C^{\pm\,\prime}$, but in fact it turns out to make little quantitative difference. 

\subsubsection{Effect of impurities on model (B)}
\label{sec:bimpure}

The mean field picture of spin-flipping processes is implicitly given in
Sec.~\ref{sec:mf-general}. Indeed, one can see that interchanging $p_\ell^+$
and $p_\ell^-$ in Eqs.~(\ref{eq:jp}),~(\ref{eq:jm}) for the mean field
current of bond $\ell$, $\ell+1$ without impurity represents the spin flip
and correct current with an impurity residing on site $\ell$. 

Recall that model (B) with unpolarized injection is unaffected by the (equivalent)
flipping of plus and minus spins, so that case is covered by the results in
Secs.~\ref{sec:bchainpure} and~\ref{sec:bcntpure}, for chains and nanotubes
respectively. We thence proceed to treat the nanotube with fully polarized
injection (the corresponding case for the chain being trivial).

In the nanotube one has the branching and recombination of particle paths, leading 
to the statistical features already discussed for model (A) in Ref.~\onlinecite{hex13}.
For model (B) the new effects, caused by the allowed conditional double occupancy,
are most apparent at high densities of particles of both spins. This is evident
for the unpolarized injection case in the low current high-density situation occurring,
e.g., for $\alpha^+=\alpha^- \gtrsim \beta^+=\beta^-$,
see the simulation results in Fig.~\ref{fig:rhovsxa75b25}. 
But in all cases, the profiles
discussed in Sec.~\ref{sec:bcntpure} are transformed by any specific configuration of
impurities into successive sections between the impurities in which the $p_\ell^+$
and $p_\ell^-$ are interchanged. These give the qualitative features seen in the
configuration-specific simulation results shown in Figs.~\ref{fig:rhovsxa5b5} 
and~\ref{fig:rhovsxa75b25}.

Quantitative comparisons are possible using the density profiles for the pure
nanotube for each successive section. Similarly, the mean field spin-up
and spin-down particle currents $J^+$, $J^-$ given previously are interchanged by
the impurities, and their values can be compared with simulation results.

\section{TASEP model: numerics}
\label{sec:tasep-num}

\subsection{Introduction}
\label{sec:num-intro}

For a structure with $N_b$ bonds, an elementary time step consists of $N_b$ sequential bond update
attempts,  each of these according to the following rules: (1) select a bond at random, say, bond $ij$,
connecting sites $i$ and $j$; 
(2) if the chosen bond has an occupied site to its left and an  empty site to its right, then
(3) move the particle across it, i.e., from $i$ to $j$ with probability (bond rate)  $p_{ij}$.
If an injection or ejection bond is chosen, step (2) is suitably modified to 
account for the particle reservoir (the corresponding bond rate being, respectively,  $\alpha$ or $\beta$). Thus, in the course of one time step, some bonds may be selected
more than once for  examination and some may not be examined at all.
This constitutes the {\it random-sequential update} procedure described
in Ref.~\onlinecite{rsss98} for the 1D TASEP, which is the realization of the usual master equation in continuous time.

In order to account for particle spin,  adaptations are needed. As in the original TASEP rules,  we stick to the interpretation that a successful bond update attempt  means  the motion of a {\em single} particle across that bond. In model (B), for step (2) above
one allows double occupancy if in agreement with Pauli's principle; furthermore,  if site $i$
itself is doubly occupied and site $j$ is empty,  then either particle may be moved from $i$
to $j$ with $50\%$ probability.   
For step (3),  with $\alpha^\uparrow$, $\alpha^\downarrow$ being the mutually exclusive, spin-dependent, particle injection rates  ($\alpha \equiv \alpha^\uparrow + \alpha^\downarrow$), 
spins are independently chosen with probability $P^{\uparrow\,(\downarrow)} =\alpha^{\uparrow\,(\downarrow)} /\alpha$ for individual injection attempts.
The case $\alpha^\uparrow=\alpha$ corresponding to a fully polarized injected current is of special interest.
Also, the case of a fully depolarized incoming current,  $\alpha^\uparrow = \alpha^\downarrow =
\alpha/2$ will be considered below. With regard to ejection,  if the rightmost site is singly
occupied  the particle is ejected with spin-independent probability $\beta$; for double
occupation, the particle to be ejected (also with probability $\beta$) is chosen with $50\%$ probability.
 The ejection procedures just delineated are consistent with our definition of an elementary
bond update as involving  crossing of the bond by at most a single particle; comments on the
connection of this to actual experimental situations are made in Sec.~\ref{sec:conc}. For the
remainder of this paper, this amounts to replacing the ejection rates $\beta^+$, $\beta^-$
introduced in Sec.~\ref{sec:mf-general}
with a single, spin-independent, parameter $\beta$.

We evaluated steady-state currents, density profiles, and (normalized) spin polarizations.
For the nanotube geometry, 
the steady-state current $J$ is the time- and ensemble-averaged number of particles
which enter  the system per unit time, divided  by the number $N_w$ of parallel entry channels (to provide proper comparison with the strictly one-dimensional case).
For each given realization of quenched randomness (collection of randomly-chosen locations of spin-flipping  sites) we repeated the following procedure $N_{\rm sam}=10$--$100$ times (each time with a distinct seed, i.e., producing $N_{\rm sam}$ independent samples
 of the stochastic update process): starting with an empty lattice, we kept injecting 
 spin-up particles  into the system's
left edge, at a fixed  injection rate $\alpha$; after waiting for a suitable time until steady-state
conditions set in, we would take $N_{\rm max} = 10^5$--$10^6$ successive realizations of stochastic update. As is well known~\cite{rbsdq12,dqrbs96}, the sample-to-sample RMS deviations for quantities estimated in this
way are essentially independent of $N_{\rm sam}$
as long as $N_{\rm sam}$ is not too small, and vary as $N_{\rm max}^{-1/2}$. Furthermore,  we considered $N_q$ independent realizations of quenched disorder;  for reasons to be explained below,
both  cases $N_q=1$ (fixed-sample) and $N_q > 1$ (typically of order $100$) turn out to be
of interest.  In contrast to the stochastic aspects just mentioned,  disorder-associated sampling 
does not produce a distribution of results whose width shrinks with growing
sample size:  the pertinent distributions display a permanent spread, as will be exemplified
in the following.  

\subsection{Chain without impurities: maximal current}
\label{sec:num-chain}

In order to test the mean field theory of Sec.~\ref{sec:bchainpure},
in particular the 
predictions  of Eqs.~(\ref{eq:jvsxast})--(\ref{eq:rsxast}) concerning equal 
level profiles and the maximal
current for model (B) on a chain with no impurities, we took unpolarized injection
with $\alpha=\beta$ ranging from $0.6$ to $1.0$.
We ran simulations with chain length $L=41$, this length  having proved sufficient 
to keep finite-size effects  contained within the error bars associated 
with intrinsic stochastic fluctuations for $N_{\rm max}=10^5$, and to provide
approximately level sections to the profiles for $\alpha=\beta \lesssim 0.75$.
For $\alpha=\beta=1$ one gets the highest total current  $J_m=0.399(7)$,
and a $\tan$-like profile consistent with the maximal current 
phase~\cite{derr98,sch00,mukamel,derr93,rbs01}.
The case $\alpha=\beta=0.75$ has $J=0.395(4)$. i.e., the same within quoted errors,
and its profiles $R$, $S$ range in the left half of the system by about $\pm 0.02$
from the value $\approx 0.47$ at the injection site.   
Thus, to a good approximation one can assume that both $(\alpha,\beta)$ pairs are within a maximal 
current phase analogous to the one exhibited by the standard TASEP. 
So the mean field prediction $J_m^-=J^+_m=0.243207 \dots$, and $R_m=0.55671 \dots$
are of the order of $20\%$ in excess of the numerical estimates.

On the other hand, the relationship Eq.~(\ref{eq:jvsalpha}), written in the form
$J^+=\alpha(1-R)$, is verified by the numerical results within stochastic error,
as should be expected because it requires no factorization 
approximation (unlike the other mean field relationships used).

\subsection{Nanotube with impurities}
\label{sec:num-cnt}

We now turn to the nanotube geometry, in presence of impurities.
In order to reduce non-essential fluctuations, we used a canonical ensemble to generate our
impurity realizations. For a nanotube with $N_s$ sites in all, and nominal impurity concentration $x_i$
we would randomly draw $m_i$ out of the $N_s$ possible locations, $m_i$ being  the integer closest to $x_i\,N_s$.  Consequently the effective concentration $x_i^{\rm eff}$ differs from $x_i$; however,
even in the physically reasonable range $x_i \ll 1$ such discrepancy is very small for the 
(relatively large) systems used. In our simulations we generally took  $x_i=0.01$; 
%%%%%%%%%%% 1st referee, point (2) %%%%%%%%%%%%%%%%%%%%%%
for  $N_w=14$, $N_r=20$ (a combination we used quite often, as seen below) one gets
$m_i=11$, giving $x_i^{\rm eff}=11/1134=0.09700 \dots$

\begin{figure}
{\centering \resizebox*{3.2in}{!}{\includegraphics*{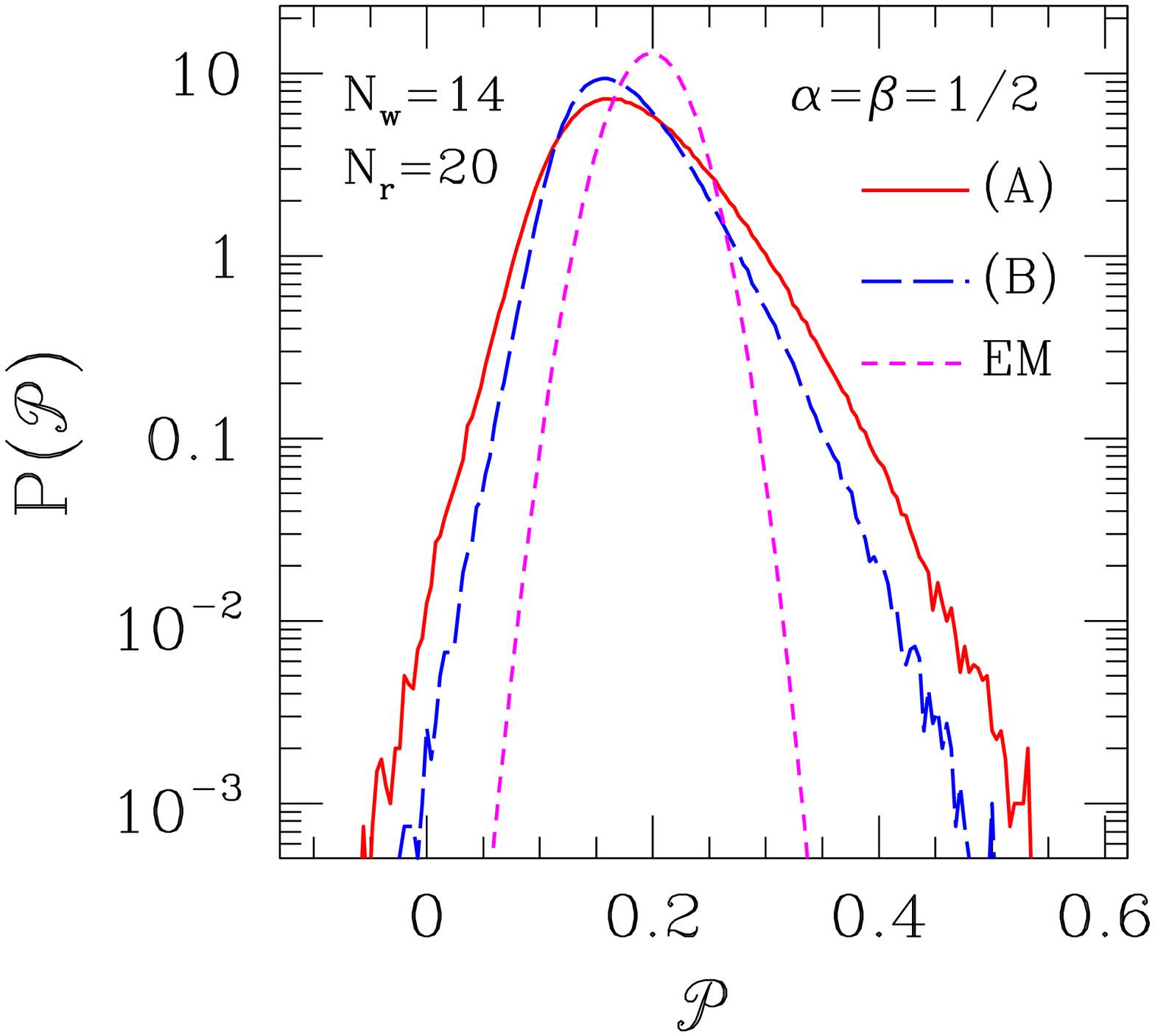}}}
\caption{%(Color online)
Log-linear plot of probability density function for polarization at the right end (exit) of a nanotube, for a current injected 
with ${\cal P}_{\rm in}=1$ at its  left end. Here $\alpha=\beta=1/2$, $N_w=14$, $N_r=20$, $x_i=0.01$.
(A) and (B) refer, respectively, to models with single or conditional-double site occupation; in both cases, samples  were taken over $N_q=10^6$ distinct realizations of impuritiy configurations.
EM refers to the effective-medium description given in Sec.~\ref{sec:tasep-theo}. 
}
\label{fig:pab05pdf}
\end{figure}

We started by fixing the external rates at $(\alpha,\beta)=(1/2,1/2)$. For a nanotube with
$N_w=14$, $N_r=20$ we examined the decay of the polarization ${\cal P}$ of a fully-polarized
current injected at the system's left end, against position along the average  flow direction.
We collected  data from $N_q=10^6$ independent samples of impurity configurations, in order
to produce smooth probability density curves at the right  (exit) end, where an ejected particle
would necessarily have gone through $81$ sites. The results for models (A) and (B), as well as for
the effective-medium (EM) approach described in Sec.~\ref{sec:tasep-theo}, are  shown
in Fig.~\ref{fig:pab05pdf}. Averages and RMS dispersions are as follows:
$\langle {\cal P}\rangle=0.184(60),\  0.177(48),\  0.199(31)$  
 respectively for (A), (B), and EM.
It is worth remarking that  already with $N_q=100$  the numerical estimates for
 both average polarizations and dispersions are very close to those just quoted: one gets
  $\langle {\cal P}\rangle=0.181(60),\  0.173(48)$ 
% CNT ../../kpz/so_hex/spc2dbpc.out, spc2dbpc_cdo.out
 respectively for (A), (B), although the corresponding distributions are of course rather spiky and
 shapeless. 

So, in this case: (i) whether single- or conditional-double occupancy is allowed has no clearly discernible 
effect on polarization decay; also (ii) the steady-state current through the system is
  $J=0.3064(1)$ in (A), $0.307(1)$ in (B),  % CNT  ../../kpz/so_hex/spcdistbpc.out, spcdistbpc_cdo.out
 identical within error bars (and in line with results for spinless systems with
the same $(\alpha,\beta)$~\cite{hex13}); furthermore, (iii) although the EM description predictably underestimates fluctuations, 
its result for the polarization distribution at the right end still falls well within the  broader dispersion of both numerically-evaluated curves.  

\begin{figure}
{\centering \resizebox*{3.2in}{!}{\includegraphics*{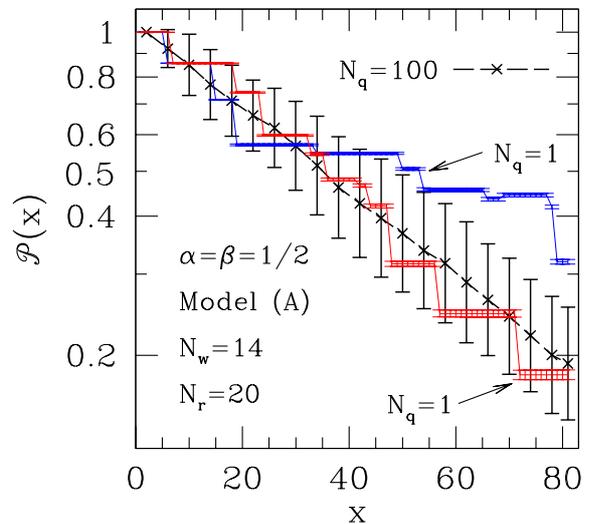}}}
\caption{%(Color online)
Log-linear plot of polarization against position $x$ along average flow direction of a nanotube, for a current injected 
with ${\cal P}_{\rm in}=1$ at its  left end. Here $\alpha=\beta=1/2$, $N_w=14$, $N_r=20$, $x_i=0.01$. Each of the polarization profiles  denoted $N_q=1$  (blue and red) corresponds to a distinct, fixed, realization of the impurity distribution.
}
\label{fig:pvsxab05}
\end{figure}

Next, we compare fixed-sample ($N_q=1$) versus multiple-sample polarization results, still at
$(\alpha,\beta)=(1/2,1/2)$. For $N_q=100$, Fig.~\ref{fig:pvsxab05} again shows  the broad scatter 
associated with sampling over disorder configurations. By contrast, for the two examples corresponding to $N_q=1$ (where the sharp downward steps correspond to $x$-values of the particular locations of
spin-flipping sites in the respective disorder realization), the amount of spread (related exclusively to
sampling over stochastic updates) is quite suppressed, as anticipated above.  

It can be seen that the central estimates for the $N_q=100$ curve are well aligned, suggesting a
simple exponential dependence, ${\cal P}(x) \propto \exp(-x/x_0)$ against position. A fit gives 
$x_0=47.9(2)$.
This is to be compared with $N_0=49.5$ from Eq.~(\ref{eq:binom2}) with 
$x_i=0.01$.
 Although the large scatter associated with each individual data point means that
 only limited significance can be attached to this result, it is remarkable that the sequence
 of average polarizations behaves so regularly.  
 
 Similar calculations for model (B) gave results very close to those displayed, for model (A), in 
Fig.~\ref{fig:pvsxab05}. 

\begin{figure}
{\centering \resizebox*{3.2in}{!}{\includegraphics*{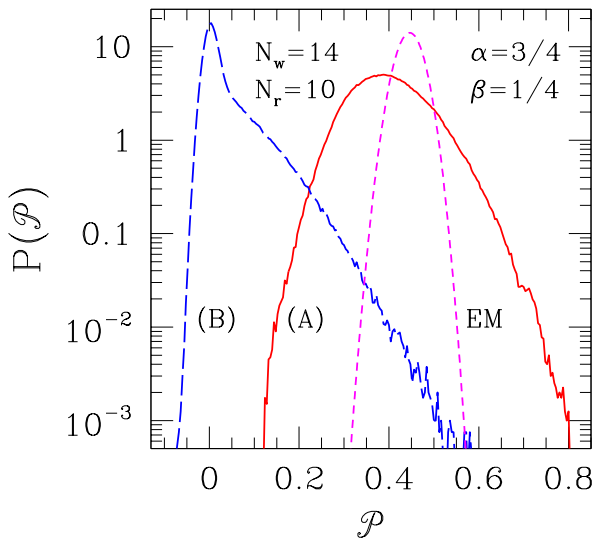}}}
\caption{%(Color online)
Log-linear plot of probability density function for polarization at the right end (exit) of a nanotube, for a current injected with ${\cal P}_{\rm in}=1$ at its  left end. Here $\alpha=3/4$, $\beta=1/4$, $N_w=14$, $N_r=10$, $x_i=0.01$.
(A) and (B) refer, respectively, to models with single or conditional-double site occupation; in both cases, samples  were taken over $N_q=10^6$ distinct realizations of impurity configurations.
EM refers to the effective-medium description given in Sec.~\ref{sec:tasep-theo}. 
}
\label{fig:pa75b25pdf}
\end{figure}

Changing the external rates to $(\alpha,\beta)=(3/4,1/4)$ produced drastically distinct results for
polarization decay, especially regarding  differences between models (A) and (B). This is seen in
Fig.~\ref{fig:pa75b25pdf}, where data for $N_w=14$, $N_r=10$ are shown.
A shorter system than for $(\alpha,\beta)=(1/2,1/2)$ was used in order to produce a nontrivial  structure of the distribution for model (B); had we used $N_r=20$ this would give essentially a delta function centered at zero, on the scale of Fig.~\ref{fig:pa75b25pdf} (see the corresponding entry
in Table~\ref{t1} below). So, while the probability density function for exiting polarization 
in model (A) compares to the EM prediction in a similar manner to the case  
$(\alpha,\beta)=(1/2,1/2)$, allowing double occupancy here has a strong polarization-curbing effect. 
 
We then varied $\alpha$ and $\beta$, probing selected points in the parameter space. Our results are displayed in Table~\ref{t1}.
 \begin{table}
\caption{\label{t1}
For systems with $x_i=0.01$, $N_w=14$,  $N_r=20$, and $\alpha$, $\beta$ as specified, 
$J(A)$, $J(B,X)$ are steady-state currents for: model (A) with initial polarization 
${\cal P}_{\rm in}=1$,
and model (B) with  ${\cal P}_{\rm in}=X$, $X=0$ or $1$.  ${\cal P}_{\rm ex}$ denotes polarization at
the system's exit. All from numerical simulations with $N_{\rm max}=10^6$, $N_q=100$.
See text for further explanation of groups (I), (II), and (III).  
}
\vskip 0.2cm
\begin{ruledtabular}
\begin{tabular}{@{}cccccc}
$\alpha,\beta$ & $J(A)$  & $J(B,1)$ & $J(B,0)$ & ${\cal P}_{\rm ex}(A)$ & ${\cal P}_{\rm ex}(B,1)$  \\
\hline\noalign{\smallskip}
\multicolumn{6}{c}{(I)}\\
\hline\noalign{\smallskip}
$1/4,1/4$ & $0.1963(1)$  &  $0.1963(1)$  & $0.2184(1)$ & $0.179(56)$ & $ 0.184(54)$  \\
$1/4,1/2$ & $0.1963(1)$  &  $ 0.1963(1)$  & $0.2184(1)$ & $0.178(54)$ & $ 0.176(50)$  \\
$1/4,3/4$ & $0.1963(1)$  &  $0.1963(1)$  & $0.2184(1)$ & $0.188(64)$ & $0.181(52)$\\
$1/2,1/2$ & $0.3064(1)$  &  $0.307(1)$  & $0.3755(1)$ & $0.183(53)$ & $0.177(47)$  \\
$1/2,3/4$ & $0.3064(1)$  &  $0.307(1)$  & $ 0.3755(1)$ & $0.179(61)$ & $0.176(44)$  \\
\hline\noalign{\smallskip}
\multicolumn{6}{c}{(II)}\\
\hline\noalign{\smallskip}
$1/2,1/4$ & $0.2133(1)$  &  $ 0.2457(1)$  & $0.2457(1)$ & $0.187(69)$ & $0.001(4)$  \\
$3/4,1/4$ & $0.2133(1)$  &  $0.2457(1)$  & $0.2456(1)$ & $0.190(68)$ & $0.000(4)$  \\
\hline\noalign{\smallskip}
\multicolumn{6}{c}{(III)}\\
\hline\noalign{\smallskip}
$3/4,1/2$ & $0.3336(1)$  &  $0.355(5)$  & $ 0.4561(2)$ & $0.184(56)$ & $ 0.168(39)$  \\
$3/4,3/4$ & $0.34817(4)$  &  $0.355(5)$  & $0.4771(1)$ & $0.177(46)$ & $ 0.177(46)$  \\
$1,1$  & $0.35069(3)$  &  $0.369(9)$  & $ 0.5314(1)$ & $0.188(64)$ & $0.168(44)$  \\
\end{tabular}
\end{ruledtabular}
\end{table}

 The main feature evinced is a clear separation into three distinct patterns of behavior.
For group (I), one has (with the definitions given in the caption to Table~\ref{t1}):
 $J(A)$ depending only on $\alpha$, $J(A)=J(B,1) < J(B,0)$, and (within  error bars) 
${\cal P}_{\rm ex}(A)={\cal P}_{\rm ex}(B,1)$. For group (II)   $J(A)$ depends only on $\beta$, $J(A) <J(B,1) = J(B,0);
 {\cal P}_{\rm ex}(B,1)=0$ while  ${\cal P}_{\rm ex}(A)  >0$. For group (III) one has $J(A) <J(B,1)<J(B,0)$
 and ${\cal P}_{\rm ex}  >0$ for both $(A)$ and $(B,1)$. Note that for all cases with ${\cal P}_{\rm ex}  >0$
 the prediction of the effective-medium approach of Sec.~\ref{sec:tasep-theo}, namely that
${\cal P}_{\rm ex}$ does not depend on $\alpha$, $\beta$,  seems to be qualitatively and,
to a reasonable extent quantitatively, fulfilled.

 As expected from the mean field theory in Sec.~\ref{sec:theo-mf}
one can draw a correspondence between group (I) and the low-current, low-density phase of 1D TASEP~\cite {derr98,derr93}   where $J$ is determined singly by $\alpha$ for $\alpha < \beta$, $\alpha <1/2$. Similarly,
group (II) has its analogue in the low-current, high density phase $\alpha > \beta$, $\beta <1/2$
of the 1D case where $J$  depends only on $\beta$. Finally, group (III) seems to be akin to the 1D maximal-current phase
at $\alpha,\beta > 1/2$, although this remark will have to be qualified, as seen below.   A phase with maximal-current features has been found for
large $\alpha$, $\beta$ in  earlier studies of TASEP on honeycomb lattices, see Fig.~7 of Ref.~\onlinecite{hex13}.

It is to be noticed that the fractional standard deviations of currents in Table~\ref{t1} are, in
general, of similar magnitude in models (A) and (B), and much smaller than those of exit polarizations (when the latter are nonzero),  even though
both quantities result from  averaging over quenched disorder configurations. Although such feature is certainly expected for model (A)  where current and polarization aspects are fully decoupled, it is not obviously forthcoming in model (B). To understand this it must be kept in mind that, for each fixed disorder configuration, current evaluation involves extensive stochastic sampling of allowed particle motions. Averaging over the latter ensemble has the effect that, in model (B) as well as (A),  the influence of spin-flipping impurities on the system-wide current shows up only through their overall concentration (as opposed to depending on their specific locations).

Further insight into the contrasting behavior of models (A) and (B) in groups (I) and (II) 
can be gained by studying the respective  steady-state density profiles.
In order to have an unbiased view of the stochastic effects involved in establishing and maintaining the stationary regime, we  suppressed  fluctuations related to sampling over quenched disorder by using the {\em same}  fixed impurity configuration ($N_q=1$) in all cases depicted in the 
following Figs.~\ref{fig:rhovsxa5b5} and ~\ref{fig:rhovsxa75b25}.

%%%%%%%%%%%%%%%%% Editors' point (1/3) %%%%%%%%%%%%%%%%%%%%%%%%%
\begin{figure}
{\centering \resizebox*{3.2in}{!}{\includegraphics*{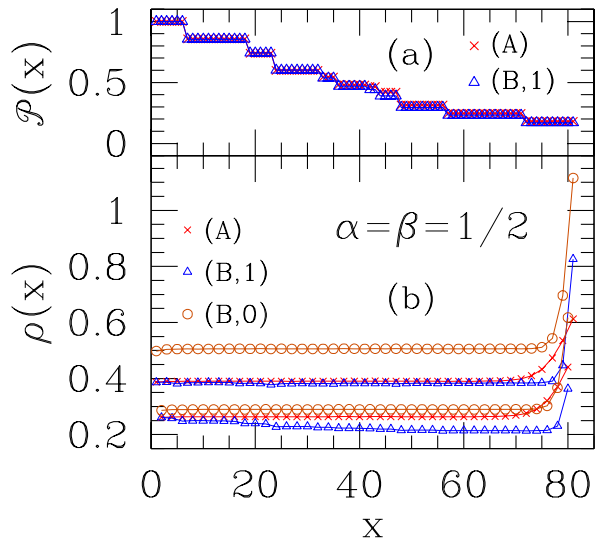}}}
\caption{%(Color online)
 (a) Polarizations ${\cal P}(x)$ and  (b) full densities $\rho(x)$ 
(spin-up plus 
spin-down)  against position $x$ along average flow direction of a nanotube
with  $N_w=14$, $N_r=20$, $x_i=0.01$, $(\alpha,\beta)=(1/2,1/2)$  for a current injected with ${\cal P}_{\rm in}=1$ $[\,A,\, (B,1)\,]$ or  ${\cal P}_{\rm in}=0$ $\,[(B,0)\,]$ at its  left end. In all cases, the same fixed impurity realization has been used ($N_q=1$);  for each of models (A) and (B), the two 
sequences of distinct sublattice densities~\cite{hex13,hex15} are plotted separately for ease of visualization. 
}
\label{fig:rhovsxa5b5}
\end{figure}

For the low-density case $(\alpha,\beta)=(1/2,1/2)$, the density profiles are very similar for both models, while the fixed-sample polarization results are nearly identical on the scale of 
Fig.~\ref{fig:rhovsxa75b25}. This confirms that  the additional degree of freedom provided by allowing double occupation  plays only a minor role here. 

 Furthermore, we recall that domain-wall theory predicts, and it  has been numerically verified~\cite{dw16}, that for ordinary TASEP with spinless particles on staggered chains the  difference between steady-state sublattice density profiles is constant, i.e., $x-$independent.
This feature turns out to hold in the case of Fig.~\ref{fig:rhovsxa5b5} for model (A) and, to a very good extent, for model (B) with ${\cal P}_{\rm in}=0$, but not so for (B) with ${\cal P}_{\rm in}=1$.
Such agreement would be expected for model (A) where spatial and spin degrees of freedom are effectively decoupled, because it is known that (i) domain-wall and mean-field  theory give
identical predictions for steady-state properties  of TASEP on (homogeneous or staggered) 
1D chains~\cite{dw16};  
and (ii) the mean field description of TASEP on a hexagonal lattice with uniform bond rates 
coincides with that of a chain with alternating bond rates $p_1=1$, $p_2=1/2$~\cite{hex13,hex15}. On the other hand, the fact that the constant-difference effect carries across to model (B) with 
${\cal P}_{\rm in}=0$, but not if  ${\cal P}_{\rm in}=1$,  indicates that the former behaves   
throughout the system similarly to  the flow of two immiscible fluid species with equal local densities. For the latter, on average the spin-flipping sites  provide transformation of the spin-up ``species''  onto the spin-down one along the system, until polarization finally approaches zero, and the sublattice density
differences approach a constant value.  So the effects of  such ``species transmutation'' are not included in the mean field approach which predicts constant density profile differences. 

We have verified that the ``immiscible species'' picture  is only semi-quantitatively correct, on account of the long-range density correlations known to exist generally in the TASEP.  Indeed,  the injection and ejection rules described above suggest that, as far as densities are concerned one could approximate the flow of an incident current
with  ${\cal P}_{\rm in}=0$  at rates $\alpha$, $\beta$ in model (B) as two separate copies of the  flow of a spinless fluid at rates $\alpha/2$, $\beta$  in model (A). However, for the system described in 
Fig.~\ref{fig:rhovsxa5b5}, the nearly constant single-spin densities for $5 \lesssim x \lesssim 70$ are
$\rho_1 \approx 0.253$, $\rho_2 \approx 0.145$ with half-current $J/2 \approx 0.1878$ 
[$\,$from Table~\ref{t1}$\,$] in (B,0)  with $(\alpha,\beta)=(1/2,1/2) $,
while in model (A) with $(\alpha,\beta)=(1/4,1/2)$   total particle densities are $\rho_1 \approx 0.215$, $\rho_2 \approx 0.127$ (not shown in the Figure)  with $J=0.1963(1)$ 
[$\,$from Table~\ref{t1}$\,$]. 

%%%%%%%%%%%%%%%% Editors' point (2/3) %%%%%%%%%%%%%%%%%%%%%%%%%%%%%%%
\begin{figure}
{\centering \resizebox*{3.2in}{!}{\includegraphics*{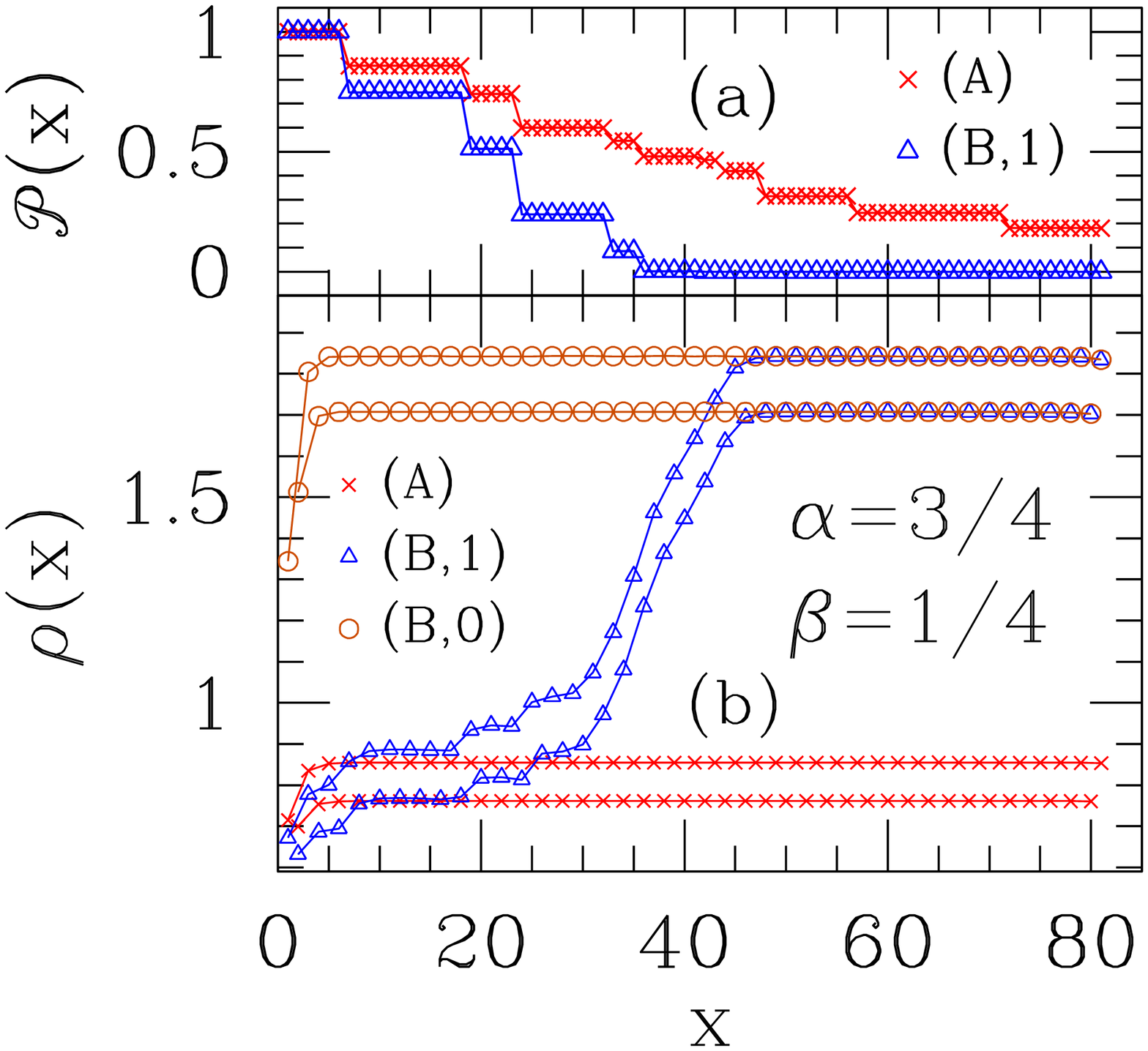}}}
\caption{%(Color online)
 (a) Polarizations ${\cal P}(x)$ and (b) full densities $\rho(x)$ 
(spin-up plus 
spin-down) against position $x$ along average flow direction of a 
nanotube with $N_w=14$, $N_r=20$, $x_i=0.01$, $(\alpha,\beta)=(3/4,1/4)$ 
for a current injected with ${\cal P}_{\rm in}=1$ $[\,A,\, (B,1)\,]$ or ${\cal 
P}_{\rm in}=0$ $\,[(B,0)\,]$ at its left end. In all cases, the same fixed
impurity realization has been used ($N_q=1$); for each of models (A) and 
(B), the two sequences of distinct sublattice 
densities~\cite{hex13,hex15} are plotted separately for ease of 
visualization.
}
\label{fig:rhovsxa75b25}
\end{figure}

In the high-density case $(\alpha,\beta)=(3/4,1/4)$ depicted in 
Fig.~\ref{fig:rhovsxa75b25},
double-occupancy is paramount for the steady-state profile configurations, especially
on the system's right half.  The fact that the current is the same in model (B), both for initial  polarization equal to 
one or zero (see Table~\ref{t1}), confirms that the low-$\beta$ bottleneck on the right plays the
dominant role in establishing stationary flow.  

Regarding the above considerations of the ``immiscible species'' picture for the flow with
${\cal P}_{\rm in}=0$, one sees in Fig.~\ref{fig:rhovsxa75b25} that 
the spatial extent $x \gtrsim 35$ of the region where ${\cal P}(x)$ has essentially
vanished is slightly longer than that, $x \gtrsim 45$, where the profiles coincide for
 ${\cal P}_{\rm in}=0$ and $1$. So, density-wise the  case (B,1) crosses over from a behavior like 
 that of (A) , up to $x \lesssim 20$, towards that of (B,0) albeit with a ``healing length'' 
 corresponding to   $35 \lesssim x \lesssim 45$ along which, although ${\cal P}=0$ already,
 the local densities have not fully converged to the same values as for (B,0).

\begin{figure}
{\centering \resizebox*{3.2in}{!}{\includegraphics*{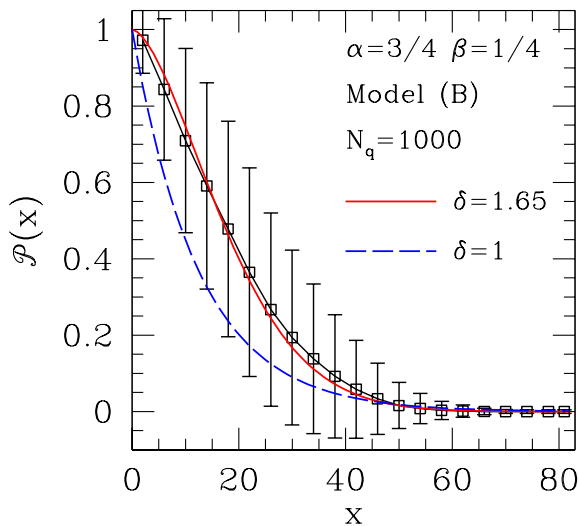}}}
\caption{%(Color online)
Polarization against position $x$ along average flow direction of a nanotube, for a current injected with ${\cal P}_{\rm in}=1$ at its  left end. Here $\alpha=3/4$,$\beta=1/4$, $N_w=14$, $N_r=20$, $x_i=0.01$. Average over $N_q=1000$ distinct realizations of the impurity distribution. The curves correspond
to fits of central estimates to a form ${\cal P}(x) = \exp \left[-(x/x_0)^\delta\right]$ (see text).
}
\label{fig:pvsxa75b25}
\end{figure}

In Fig.~\ref{fig:pvsxa75b25} we show polarization against position (averaged over quenched disorder
realizations) for model (B) with   $(\alpha,\beta)=(3/4,1/4)$. It is clear that, contrary to the case
depicted in Fig.~\ref{fig:pvsxab05}, fitting a pure exponential form to the sequence of central estimates (blue, long-dashed curve in  Fig.~\ref{fig:pvsxa75b25}) gives unsatisfactory results. However,  considering a generalized exponential function  
${\cal P}(x) = \exp \left[-(x/x_0)^\delta\right]$
gives a much closer fit with $\delta=1.65(4)$; see the full red line in  Fig.~\ref{fig:pvsxa75b25}.

We investigated the effects of system size on the results exhibited so far.  In agreement  with
previous studies for spinless sytems~\cite{hex13,hex15} the dependence of overall currents
and average densities on transverse ($N_w$) and longitudinal ($N_r$) dimensions is rather weak.  
Regarding polarization-specific features, we checked the characteristic decay lengths $x_0$,
both in the low- and high-density phases, as well as the phenomenological parameter $\delta$
for the latter case. As expected, the $N_w$ dependence is residual. We found also that $x_0$
as well as $\delta$ (the latter,  where pertinent) also exhibit only a weak $N_r$ dependence.
This is not obvious from the outset, especially for the high-density phase in model (B), given the 
influence of low $\beta$ at the ejection sites on the buildup of double occupation backwards from 
there (see Figs.~\ref{fig:rhovsxa75b25} and~ \ref{fig:pvsxa75b25}); in this case, for $N_r=20,$ $30$, $40$ one gets [$\,$keeping $\delta=1.65$ fixed$\,$] $x_0=21.1(4)$,  $23.8(7)$, and  $25.7(9)$ respectively.

\begin{figure}
%%%%%%%%%% Editors' point  (3/3) %%%%%%%%%%%%%%%%%%%%%%%%%%%
{\centering \resizebox*{3.2in}{!}{\includegraphics*{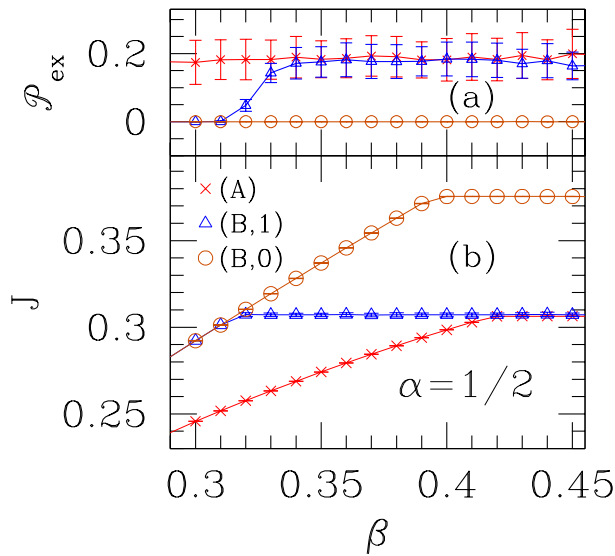}}}
\caption{%(Color online)
 (a) Polarization at right end  and  (b) steady-state current $J$  for a 
nanotube with $N_w=14$, $N_r=20$, $x_i=0.01$, for a current injected with  
${\cal P}_{\rm in}=1$ $[\,A,\, (B,1)\,]$ or ${\cal  P}_{\rm in}=0$ $\,[(B,0)\,]$ at its left end, against ejection rate $\beta$.   Injection rate $\alpha=1/2$ (fixed). Averages over $N_q=100$ distinct realizations of the impurity distribution. 
}
\label{fig:a05bvar}
\end{figure}

Next we examined the transition between the patterns of behavior characterizing groups (I) and (II)
of Table~\ref{t1}. Keeping  $\alpha=1/2$ fixed we varied $\beta$ from  $0.25$,
within the high-density (HD) phase, to $0.50$, within the low-density (LD) one. Fig.~\ref{fig:a05bvar}
shows exit polarization ${\cal P}_{\rm ex}$ and steady-state current $J$.
One sees that for $\beta \approx 0.32$ the current for (B,1) sharply turns from being equal to
that of (B,0)  to a constant value against increasing $\beta$, indicating an $\alpha-$dominated 
regime there, and eventually merging with $J(A)$ for $\beta \gtrsim 0.43$.  Though the departure
in  behavior of ${\cal P}_{\rm ex}$ for (B,1) from the (B,0) pattern is not as clearly demarked as for 
$J$, it converges faster to that of (A), say by $\beta \approx 0.34$. We scanned the $\beta$ axis
also for  different values of fixed $\alpha$, namely $1/4$, $3/4$, and $1$. In all cases we found
the same qualitative picture as that given for $\alpha=1/2$ in Fig.~\ref{fig:a05bvar}.
The $\alpha-$ dependence of the $\beta$ value for which the current for (B,1) becomes constant against varying  $\beta$ is given approximately by
\begin{equation}
 \beta(\alpha)=0.8\alpha-0.4\alpha^2\ .
 \label{eq:beta(alpha)}
 \end{equation}
Going back to Table~\ref{t1} and referring to Eq.~(\ref{eq:beta(alpha)}), one 
sees that the (B,1) results for $(\alpha,\beta)=(3/4,3/4)$ and $(3/4,1/2)$ are 
both associated with the intermediate-behavior section analogous to the $0.32 
\lesssim \beta \lesssim 0.43$ stretch in Fig.~\ref{fig:a05bvar}.

Corresponding analysis of numerical data for the unpolarized case (B,0)
provides a test of the theoretical results obtained in Eq.(\ref{eq:beta(a)_th}), 
see Table~\ref{t2}. One sees that the agreement between mean field theory and
numerics is very good in this case.

\begin{table}
\caption{\label{t2} For model (B,0) and selected values of $\alpha$,
$\beta_{\rm th}$ are the values of $\beta$ for which total
current $J$ becomes $\beta$-independent, as predicted by Eq.(\ref{eq:beta(a)_th}); 
$\beta_{\rm num}$ are
numerically-obtained results (see, e.g., Fig.~\ref{fig:a05bvar}).
}
\begin{ruledtabular}
\begin{tabular}{@{}ccc}
$\alpha$ & $\beta_{\rm th}$  & $\beta_{\rm num}$  \\
\hline\noalign{\smallskip}
$1/4$ & $2/9$  &  $0.23(1)$  \\
$1/2$ & $2/5$  &  $0.40(1)$  \\
$3/4$ & $6/11$  &  $0.55(2)$  \\
$1$ & $2/3$  &  $0.65(2)$  \\
\end{tabular}
\end{ruledtabular}
\end{table}

\begin{figure}
{\centering \resizebox*{3.2in}{!}{\includegraphics*{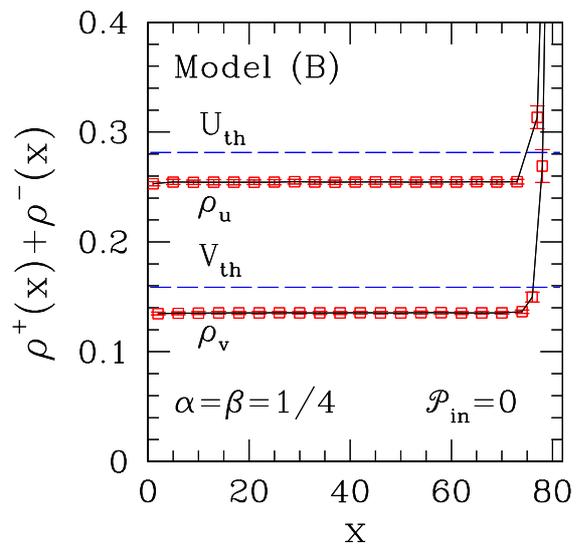}}}
\caption{%(Color online)
Points ($\rho_u$ and $\rho_v$) show steady-state sublattice full (spin-up  plus spin-down) densities against  position along flow direction, for model (B) on a nanotube
with $N_w=14$, $N_r=20$ at $(\alpha,\beta)=(1/4,1/4)$. ${\cal P}_{\rm in}=0$, so 
impurities are statistically irrelevant in establishing total current and density 
profiles. Horizontal dashed lines ($U_{\rm th}$ and $V_{\rm th}$)
 show mean field predictions derived from Eqs.~(\ref{eq:cntlevel})--(\ref{eq:cntmap}). See text.
}
\label{fig:mfpol0}
\end{figure}

In order to provide further checks of the mean field theoretical predictions from Sec.~\ref{sec:bcntpure}
we took $(\alpha,\beta)=(1/4,1/4)$.  These rates, as can be seen from Table~\ref{t1}, correspond
to a point deep in the low-current, low-density region of the phase diagram where both sublattice
density profiles are expected to be level throughout the system (apart from an upturn close to
the ejection end, see Fig.~\ref{fig:rhovsxa5b5} for a less extreme example). Also, we took
an unpolarized injected current, ${\cal P}_{\rm in}=0$ so impurities would have  no net effect.
In such conditions one would expect the conditions expressed in 
Eqs.~(\ref{eq:cntlevel})--(\ref{eq:cntmap}) to apply.  Theoretical and simulational results are
displayed, for both sublattices, in Fig.~\ref{fig:mfpol0}. The numerically-evaluated full
(spin-up plus spin-down) densities for the level section of the profiles are: 
$\rho_u=0.254(2)$, $\rho_v=0.153(2)$, to be compared to the predictions from theory. 
The latter are found by plugging the full (numerically-evaluated) steady-state 
current $J=0.2184(1)$ from Table~\ref{t1} into the specific forms of 
Eqs.~(\ref{eq:jp}),~(\ref{eq:jm}) given by Eq.~(\ref{eq:cntmap}), and solving for 
$U \equiv u(1+u)$, $V \equiv v(1+v)$. One gets $U_{\rm th}=0.2814(2)$, $V_{\rm th} 
=0.1586(1)$, respectively $10\%$ and $4\%$ in excess of simulation results.
Then inserting $J$ and $U_{\rm th}$ into left- and right-hand sides of 
Eq.~(\ref{eq:jp0}) gives $\alpha=0.3039(4)$, different by $\sim 20\%$ from 
the actual value $1/4$.

We have also checked how the characteristic decay length $x_0$ depends on impurity concentration.  In this case we restricted ourselves to $(\alpha,\beta)=(1/2,1/2)$, well within
the portion of phase space for which polarization decay follows a simple
exponential form. We considered  $x_i=0.003$, $0.005$, $0.0075$, and $0.01$, all in
the very low impurity-density regime. For model (B) with ${\cal P}_{\rm in}=1$ we took systems of
 fixed width $N_w=14$ and varying lengths in the range $20 \leq N_r \leq 120$. For fixed $x_i$
the final estimates of $x_0$ took into account the dispersion among the results of the individual fits
for each $N_r$.  The sequence of the $\{ x_0\}$ was adjusted to a power-law form, 
$x_0 \propto x_i^{-a}$, whence $a=1.03(2)$. This agrees well with the result $a=1.1(1)$
of  Ref.~\onlinecite{dq15}.
% (see  Fig. 5, and discussion around Eq.~(10), of  that Reference).

\begin{figure}
{\centering \resizebox*{3.2in}{!}{\includegraphics*{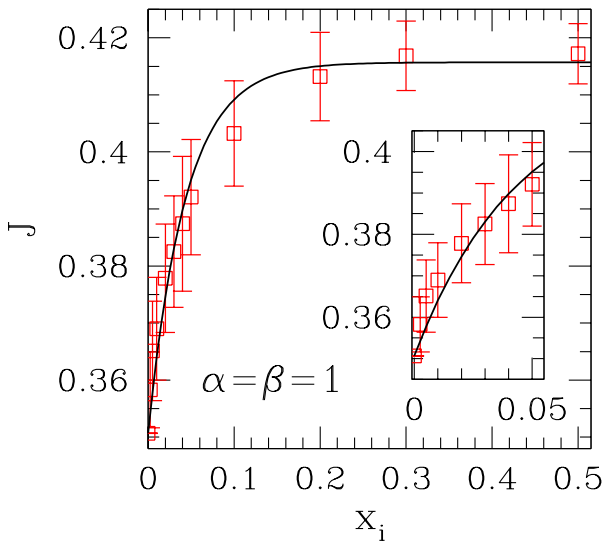}}}
\caption{%(Color online)
Steady-state current $J$ as a function of impurity concentration $x_i$ for model (B) at $(\alpha,\beta)=(1,1)$ with a fully polarized current injected at the left end 
(${\cal P}_{\rm in}=1$).
$N_w=14$, $N_r=20$. The full line is an {\em ad hoc} exponential fit to the data. 
The inset shows details of the main figure close to the vertical axis.
}
\label{fig:ab1xvar}
\end{figure}

We now illustrate how the interplay of double occupancy and spin-flipping impurities can result
in an enhancement of the steady state current across the system. Taking model (B) with a
fully polarized current injected at the left end (${\cal P}_{\rm in}=1$), we considered
$\alpha=\beta=1$. In this way the constraints imposed by boundary conditions are minimized,
and the remaining impediments to particle flow are only those associated with exclusion 
according to Pauli's principle. In order to probe asymptotic trends, we allowed the impurity
concentration to vary well beyond the physically reasonable regime $x_i \ll 1$.
The results are shown for a system with $N_r=14$, $N_r=20$ in Fig.~\ref{fig:ab1xvar}, together
with an {\em ad hoc} exponential fit to the data which gives the limiting current $J_{\rm lim}=0.416(2)$.
This is significantly higher than the value $J=0.35069(3)$ for $x_i \equiv 0$, 
the latter coinciding with the (impurity-independent) current for model (A), see Table~\ref{t1}. 
However, the impurity-induced current enhancement is not enough to equal the effect of
injecting a fully depolarized beam for model (B), in which case the (also impurity-independent) corresponding value is $J=0.5314(1)$, again from Table ~\ref{t1}.
Overall, the resulting  picture emphasizes the role played by
Pauli's principle, in creating a bottleneck near the left end of 
the system for a polarized injected beam.

\section{Discussion and Conclusions} 
\label{sec:conc}

Model (B) introduced here bears a number of similarities with those generally 
known as "two-lane TASEP" models, in particular the implementation of 
Reichenbach {\em et al}~\cite{rff06,rff07}. The latter authors consider strictly 
1D systems with conditional double-occupancy rules obeying Pauli's principle.
Their spin-flipping mechanism is purely stochastic, and it is shown that several 
nontrivial effects take place for specific (mesoscopic) ranges of its occurrence 
rate. We now outline relevant differences between our approach and theirs. 
In the present case, spin-flipping results from a combination of quenched 
and stochastic factors. These are respectively: fixed locations of spin-flipping
sites (for each given realization of impurity distribution),  and the fact that, for the 
honeycomb geometry, the paths effectively followed by particles are dynamically
and randomly chosen within a highly degenerate sample space. These features stem 
naturally from those of the physical systems under consideration here, within the 
limitations of the classical model used for their
description. For the same reason, we use distinct injection rates 
$\alpha^\uparrow$, $\alpha^\downarrow$ and a single, spin-independent
ejection attempt rate $\beta$ (in contrast with $\beta^\uparrow$, $\beta^\downarrow$
of Refs.~\onlinecite{rff06},~\onlinecite{rff07}). This corresponds to an 
experimental arrangement
where full control can be exerted, e.g., by spin filtering,  on the polarization of an incoming electron beam, but the ejection mechanism at the system's end is a voltage-based, spin-independent one. 

Model (A) discussed above turns out to be  a convenient testing ground for the 
introduction of polarization features, and also to provide useful comparisons with model
(B), especially as regards the relevance, or not, of double site occupancy in the latter. It can
be seen, e.g., in Table ~\ref{t1} and Fig.~\ref{fig:a05bvar}, that the current in model (A)
is a lower bound for that in  the (B,1) implementation, i.e., model (B) with injected polarization
${\cal P}_{\rm in}=1$ [$\,$the upper bound being given by (B,0)$\,$].  Accordingly, 
having $J(A)=J(B,1)$ for a given $(\alpha,\beta)$ correlates well with having very similar, 
though not identical, density profiles; see Fig.~\ref{fig:rhovsxa5b5}. Conversely, $J(A) < J(B,1)$
signals  marked differences in such quantities, to the extent that double occupancy is 
very frequent along  large sections of the system for the latter case, see Fig.~\ref{fig:rhovsxa75b25}.

As expected on general grounds, and from the mean field arguments in 
Sec.~\ref{sec:theo-mf}, both models (A) and (B) share with 1D TASEP the basic feature of
exhibiting regions of the $(\alpha,\beta)$ plane where the overall current is solely determined either
by injection ($\alpha$) or ejection ($\beta$) rates, and which are associated respectively with
low- or high- density profiles, see Figs.~\ref{fig:rhovsxa5b5} and~\ref{fig:rhovsxa75b25}. 
For (B,1) the dividing line  is given approximately 
by Eq.~(\ref{eq:beta(alpha)}), to be compared with the corrresponding condition
for 1D TASEP~\cite{derr98,sch00,mukamel,derr93,rbs01,be07,cmz11}, namely  $\alpha=\beta$.
Furthermore, the discussion in the preceding paragraph indicates that for (B,1) the low- (high-)
density phase is strongly connected with low (high) probabiliity of average occurrence of 
doubly-occupied sites.    

One may propose a semi-quantitative correspondence of the TASEP rates $(\alpha, \beta)$ with
physical parameters of electron transport on graphenelike structures. 
While the externally-imposed potential difference between injection and ejection points
is the qualitative analogue of the directionality imposed by TASEP rules, its low or 
high intensity
may be roughly mapped onto combinations of  $(\alpha, \beta)$  which favor low or high currents,
respectively. Additionally,  the chemical potential difference $\Delta \mu_{L,R}$ between the
graphenelike structure and leads on left and right is expected to be akin to
$\alpha$ and $\beta$, respectively. Since usually one has 
$\Delta \mu_L=\Delta \mu_R$~\cite{politou15}, this would mean that experimental setups
correspond to $\alpha=\beta$. In this case the results found here, especially  the shape of the
dividing line given by Eq.~(\ref{eq:beta(alpha)}) and consequent implications, would indicate
that double occupancy does not play a quantitatively significant role in electronic transport
on graphenelike structures.

\begin{acknowledgments}
S.L.A.d.Q. thanks Belita Koiller for suggesting the problem,  M.A.G.~Cunha for clever
insights during the early stage of calculations, A.R.~Rocha for interesting 
discussions and for pointing out relevant references, and the 
Rudolf Peierls Centre for Theoretical Physics, Oxford, for
hospitality during his visit.
The research of S.L.A.d.Q. is supported by the Brazilian agencies
Conselho Nacional de Desenvolvimento Cient\'\i fico e
Tecnol\'ogico  (Grant No. 303891/2013-0)
and Funda\c c\~ao de Amparo \`a Pesquisa do Estado do Rio
de Janeiro (Grants Nos. E-26/102.760/2012, E-26/110.734/2012, and 
E-26/102.348/2013).
\end{acknowledgments}

\end{document}